\documentclass[%
    reprint,
    superscriptaddress,
    nofootinbib,
    amsmath,amssymb,
    aps,
    prd,
]{revtex4-1}

\usepackage[export]{adjustbox}
\usepackage{graphicx}%
\usepackage[caption=false]{subfig}
\usepackage{dcolumn}%
\usepackage{lipsum}
\usepackage{booktabs}
\usepackage{bm}%
\usepackage[hidelinks]{hyperref}%
\usepackage[nameinlink]{cleveref}
\usepackage{xargs} %

\usepackage{xspace}

\usepackage[dvipsnames]{xcolor}
\usepackage{verbatim}

\usepackage{float}
\makeatletter
\let\newfloat\newfloat@ltx
\makeatother

\usepackage{algorithm}
\usepackage{algpseudocode}

\usepackage{txfonts}

\usepackage{url}

\definecolor{linkcolor}{HTML}{0b5394}
\hypersetup{
    colorlinks=true,
    linkcolor=linkcolor,     %
    citecolor=linkcolor,    %
    urlcolor=linkcolor       %
}

\newcommandx{\jb}[1][]{\textcolor{blue}{JB: #1}}
\newcommandx{\commentgregor}[1][]{\textcolor{violet}{GK: #1}}
\newcommandx{\ah}[1][]{\textcolor{magenta}{AH: #1}}
\newcommandx{\commentnikol}[1][]{\textcolor{orange}{NM: #1}}
\newcommandx{\commentian}[1][]{\textcolor{green}{IP: #1}}
\newcommandx{\commentdavid}[1][]{\textcolor{red}{DS: #1}}

\newcommand{\oja}{\mbox{OmniJet-$\alpha$}}

\newcommandx{\todocite}[1][]{\textcolor{purple}{[\textbf{CITE:} #1]}}

\AddToHook{cmd/appendix/before}{\crefalias{section}{appendix}}

\begin{document}

\title{Enhancing next token prediction based pre-training for jet foundation models}

\author{Joschka Birk}
\email{joschka.birk@uni-hamburg.de}
\thanks{Corresponding author}
\affiliation{Institut f\"{u}r Experimentalphysik, Universit\"{a}t Hamburg, 22761 Hamburg, Germany}

\author{Anna Hallin}
\email{anna.hallin@uni-hamburg.de}
\affiliation{Institut f\"{u}r Experimentalphysik, Universit\"{a}t Hamburg, 22761 Hamburg, Germany}

\author{Gregor Kasieczka}
\email{gregor.kasieczka@uni-hamburg.de}
\affiliation{Institut f\"{u}r Experimentalphysik, Universit\"{a}t Hamburg, 22761 Hamburg, Germany}

\author{Nikol Madzharova}
\email{nikol.madzharova@studium.uni-hamburg.de}
\affiliation{Institut f\"{u}r Experimentalphysik, Universit\"{a}t Hamburg, 22761 Hamburg, Germany}

\author{Ian Pang}
\email{ian.pang@physics.rutgers.edu}
\affiliation{NHETC, Dept.\ of Physics and Astronomy, Rutgers University, Piscataway, NJ 08854, USA}

\author{David Shih}
\email{shih@physics.rutgers.edu}
\affiliation{NHETC, Dept.\ of Physics and Astronomy, Rutgers University, Piscataway, NJ 08854, USA}

\begin{abstract}

Next token prediction is an attractive pre-training task for jet foundation models, in that it is simulation free and enables excellent generative capabilities that can transfer across datasets. Here we study multiple improvements to next token prediction, building 
on the initial work of OmniJet-$\alpha$.
Instead of tokenizing particles and subsequently only using the token-ID as the model input for both the generative and the classification task, we adopt a hybrid setup, which allows us to use continuous feature vectors as model input while only using token-IDs in the next token prediction target.
Secondly, we explore a combined pre-training strategy that combines masked particle modeling and generative learning objectives. Taken together, these changes greatly improve the performance in downstream classification tasks without any loss in generative performance.

\end{abstract}

\maketitle

\section{Introduction}

Foundation models~\cite{bommasani2022opportunities} applied to high energy physics (HEP) data have 
recently received substantial interest, motivated by their ability to generalize to previously unseen datasets and tasks.
Jet physics in particular has been a key target for the development of HEP foundation models (see \cite{Hallin:2025ywf} for a recent review), due to its wide practical relevance across LHC and future collider experiments,
and its well-established role as benchmark for machine learning (ML)
algorithms~\cite{Kasieczka:2019dbj}. 

Much of the activity so far in the development of foundation models for jets has centered around exploring different approaches to pre-training and fine-tuning in order to optimize the expressivity of the learned representations and the performance on downstream tasks. For example, there have been many different proposals for the pre-training objective. These range from predicting a missing entry, based either on
masking~\cite{Golling:2024abg,Leigh:2024ked,Katel:2024ygn,Wildridge:2024yeg,Bardhan:2025icr}
or next token prediction~\cite{Birk:2024knn}, to classification \cite{Qu:2022mxj,Mikuni:2024qsr,Mikuni:2025tar,Bhimji:2025isp,Ho:2024qyf}, data augmentation and contrastive
learning~\cite{Harris:2024sra, hao2025rinorenormalizationgroupinvariance} or a combination of tasks
\cite{Mikuni:2024qsr,Mikuni:2025tar,Bhimji:2025isp}. Despite all these efforts, it remains unclear to what extent different pre-training methods produce representations that align well with the demands of downstream tasks.
Different pre-training objectives correlate strongly with performance on downstream tasks: for instance some approaches enable classification but not generation or vice versa. This is clearly an unsatisfactory situation for the HEP foundation model paradigm.

Another dividing line is the reliance on ground-truth information in the training phase, which determines whether they can in-principle be trained using data or whether simulation is required. While simulation-based approaches~\cite{Qu:2022mxj,Vigl:2024lat,Mikuni:2024qsr,Mikuni:2025tar,Bhimji:2025isp,Ho:2024qyf,Harris:2024sra} can leverage supervised training on high-fidelity simulations, simulation-free approaches~\cite{Golling:2024abg,Leigh:2024ked,Birk:2024knn,Katel:2024ygn,Wildridge:2024yeg,Amram:2024fjg,Bardhan:2025icr,hao2025rinorenormalizationgroupinvariance} have the potential to be trained directly on the copious amounts of real LHC data.

This work focuses on \oja{}~\cite{Birk:2024knn} as the only model with demonstrated generative capabilities and transfer learning between tasks that can be fully trained on data.
However, these advantages so far came at a price, namely a sub-optimal performance when considering  downstream classification tasks.
Here, we identify and improve upon two limitations of the original \oja{} approach, stemming from it being an autoregressive GPT-like model.
In both cases, we modify the architecture to align more closely with the nature of our physics data, while maintaining the attractive simulation-free and generation-capable properties.

First, we show that using tokenized input during the classifier training harms classification ability. 
The main reason for tokenization is that it improves generative pre-training and generative downstream performance.
However, due to information loss and possible artifacts~\cite{Pang:2025lbs}, tokenization might not be  the most optimal choice. 
To circumvent these issues,  we
present an improved hybrid setup for next token prediction based pre-training,
where the input feature vectors are continuous while the pre-training targets
remain token-IDs. Allowing for continuous input significantly increases the
classification capabilities of our model, while keeping the token-IDs as targets
for the pre-training preserves the generative performance. 

Second, we revisit the pre-training task. It is known from natural language
processing (NLP) that models based on causal attention and next token prediction
excel at generation, whereas bi-directional approaches using masking produce
rich contextual embeddings. Recent works that have focused on combining the two
training objectives
include~\cite{lv2024analysismitigationreversalcurse,behnamghader2024llm2veclargelanguagemodels,charpentier2024gptbertboth,yu2024antlmbridgingcausalmasked,liu2023betterfewshotfinetuningperformance,gisserotboukhlef2025pretrainencodersmaskedlanguage,izdebski2025synergisticbenefitsjointmolecule}.
While \oja{} has strong generative capabilities due to its next token prediction
(NTP) pre-training, we hypothesize that tasks such as jet classification require
a contextual understanding of the jet constituents which may be better served by
bi-directional pre-training like masked particle modeling (MPM)~\cite{Golling:2024abg,Leigh:2024ked} and other methods based on masked prediction~\cite{Katel:2024ygn,Wildridge:2024yeg,Bardhan:2025icr}. 
Inspired by the results from the NLP literature, we investigate the performance
of both NTP and MPM during pre-training and subsequent fine-tuning to
classification, including a setup that combines both training objectives.
We show how pre-training on both next and masked token prediction
simultaneously leads to a drastic improvement of downstream classification performance,
while maintaining the generative capabilities.

The rest of this paper adheres to the following structure:
\Cref{sec:setup} describes the used datasets, reviews the original \oja{} method and its limitations, and details the modifications we introduce to address them. \Cref{sec:results} reports the performance
gains from (i) adapting to a continuous model input
and (ii) updating the pre-training objective to combine next token
and masked token prediction.
Our conclusions are presented in \Cref{sec:conclusion}. Additional details, analyses and ablation studies are provided in
\Cref{sec:model_architecture,sec:appendix_extended_feature_set,sec:appendix_VQVAE_performance,sec:fixed_backbone}.

\begin{figure*}
    \centering
    \includegraphics[width=0.98\textwidth]{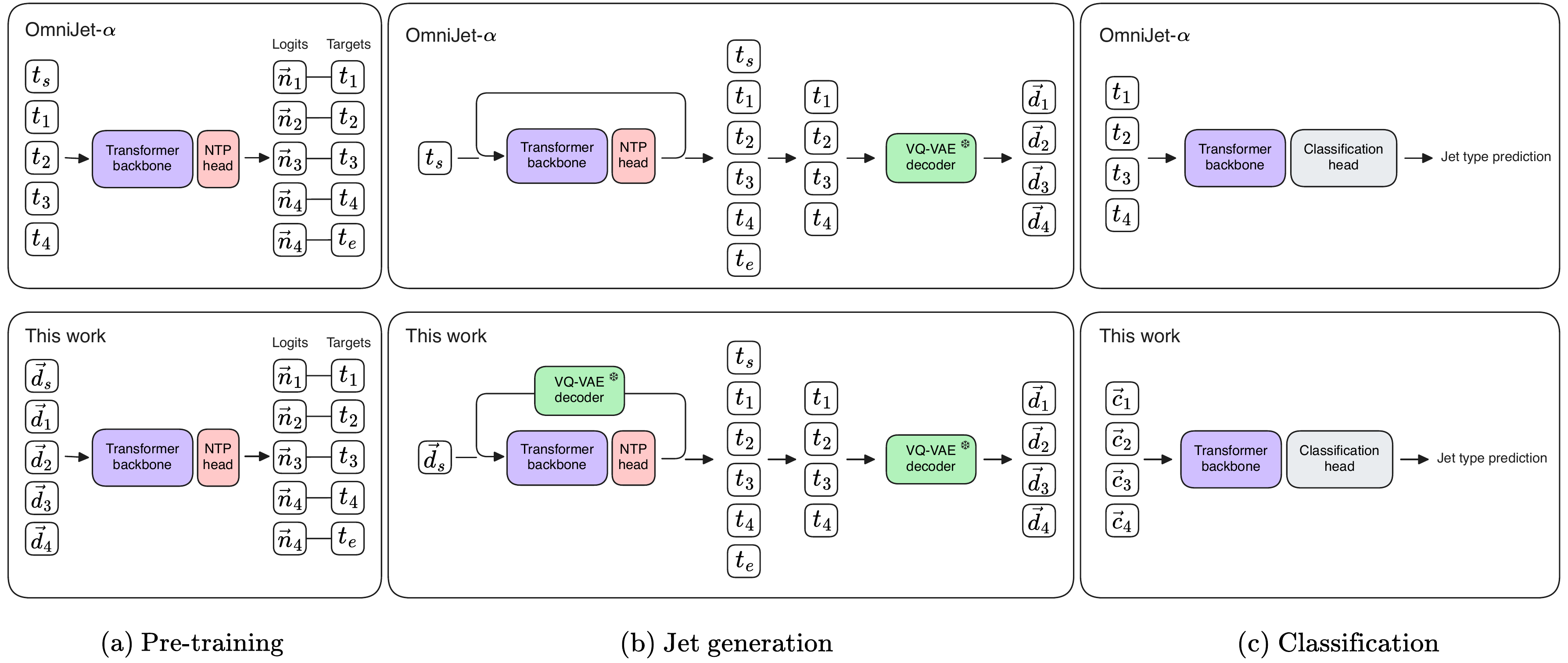}%
    \caption{
        The high-level architecture of the OmniJet workflow: (a)
        the pre-training based on next token prediction, (b)
        the generative model obtained from unsupervised pre-training and (c) the
        classification model obtained from supervised fine-tuning.
        Our new approach uses continuous feature vectors as input, both for the
        generative and classification tasks.
        Token-IDs are shown as $t_i$, with $t_s$ corresponding to a start token
        and $t_e$ corresponding to an end token.
        The continuous feature vectors are shown as $\vec{c}_i$, and the
        pseudo-continuous feature vectors (i.e. the decoded token-IDs) are shown
        as $\vec{d}_i$. In the continuous input case, the start token is a
        trainable embedding $\vec{d}_s$.
    }
    \label{fig:omnijet_alpha_workflow}
\end{figure*}

\section{Setup}
\label{sec:setup}

The model is implemented in \textsc{PyTorch}~\cite{NEURIPS2019_9015} and
\textsc{Lightning}~\cite{Falcon_PyTorch_Lightning_2019}.
Furthermore, we use the open source software libraries in Refs.~\cite{
    Hunter:2007,
    reback2020pandas,
    mckinney-proc-scipy-2010,
    harris2020array,
    awkward_paper,
    pivarski_2025_17068725,
    aryan_roy_2025_15931991,
    Cacciari:2011ma,
    huh2023vqtorch,
    scikit-learn,
    Buss_RangerLite_2025,
    Chopra2025%
}.
The overall architecture of the model is similar to the one used in \oja{},
with a few minor updates outlined in \Cref{sec:model_architecture}.

\subsection{Datasets}

Two datasets are used in our studies: the JetClass~\cite{Qu:2022mxj,qu_2022_6619768}
dataset and the top tagging dataset~\cite{kasieczka_2019_2603256,Kasieczka:2019dbj}.

The JetClass dataset contains 100\,M training jets from simulated proton-proton
collisions at the LHC.
It contains ten different jet types, with the individual jet types corresponding
to one background jet type (quark- or gluon-initiated jets) and nine signal jet
types where the jet originates from the decay of a top quark, a $W$, $Z$ or Higgs boson.
The individual jet types are:
\mbox{$q/g$},
\mbox{$t \to bqq'$},
\mbox{$t \to b\ell\nu$},
\mbox{$W \to qq'$},
\mbox{$Z \to q\bar{q}$},
\mbox{$H \to b\bar{b}$},
\mbox{$H \to c\bar{c}$},
\mbox{$H \to gg$},
\mbox{$H \to 4q$},
\mbox{$H \to \ell\nu qq'$}.
The events are simulated with MadGraph5\_aMC@NLO~\cite{Alwall:2014hca}.
Pythia8~\cite{Sjostrand:2007gs,Sjostrand:2014zea} is used for parton showering and hadronization,
and a simplified detector simulation is performed with
Delphes~\cite{deFavereau:2013fsa} using the CMS~\cite{CMS:2008xjf} detector card.
The particles are clustered with the \mbox{anti-$k_\mathrm{T}$}~\cite{Cacciari:2008gp} algorithm with
a distance parameter of $R=0.8$, and only jets with a transverse momentum in the
range of \mbox{$500\,\text{GeV} < p_\mathrm{T} < 1000\,\text{GeV}$} and a pseudorapidity
of \mbox{$|\eta| < 2.0$} are included in the dataset. 
The JetClass dataset is a widely used, public benchmark dataset for studies in
the domain of pre-training methods for jet physics. The $p_\mathrm{T}$ range used in
JetClass corresponds to boosted jets, where heavy particles ($t$, $W$, $Z$, $H$)
are fully contained within the jet.
Further details on the JetClass
dataset can be found in~\cite{Qu:2022mxj}.

The top tagging dataset contains $q/g$ jets as well as \mbox{$t \to bqq'$} jets.
This dataset is significantly smaller than the JetClass dataset, with
1.2\,M training jets. It contains jets extracted from LHC events that were
simulated with Pythia8~\cite{Sjostrand:2007gs,Sjostrand:2014zea} and
Delphes~\cite{deFavereau:2013fsa} using the ATLAS~\cite{ATLAS:2008xda} detector
card.
The jets are clustered with the
\mbox{anti-$k_\mathrm{T}$}~\cite{Cacciari:2008gp} algorithm with a distance parameter of
$R=0.8$, and are required to have \mbox{$|\eta| < 2.0$} and a transverse
momentum in the range of \mbox{$550\,\text{GeV} < p_\mathrm{T} < 650\,\text{GeV}$}.
While the $p_\mathrm{T}$-ranges of the two datasets used here overlap,
earlier work such as~\cite{Amram:2024fjg} showed successful transfer learning
across datasets with different $p_\mathrm{T}$ ranges.

\subsection{Original \oja{} method}

An overview of the method of the original work is shown in \Cref{fig:omnijet_alpha_workflow} (upper row).
In the original \oja{} work jet constituents are first
tokenized with a vector quantized autoencoder
(VQ-VAE)~\cite{Golling:2024abg,NIPS2017_7a98af17,pmlr-v202-huh23a}
to obtain a discrete representation of the continuous input features,
in the following referred to as \emph{token-IDs}, denoted as $t_i$.
A transformer architecture is used for both the encoder and decoder of
the VQ-VAE, which means that both the encoding and the decoding of
a particle is conditioned on the other particles present in the sequence.
Thus, each particle is mapped to a single token-ID, but the other particles
present in the sequence can influence the tokenization process. This allows
the tokenizer to cover the physical space more accurately as opposed to
unconditional tokenization. The VQ-VAE is trained as an individual step
before any pre-training is performed, and is frozen during the pre-trainings.
The pre-training is performed on the NTP task.
Afterwards, the integer number token-IDs are used as input for both the
generative and classification tasks. 

\begin{figure*}
    \centering
    \includegraphics[width=0.85\textwidth]{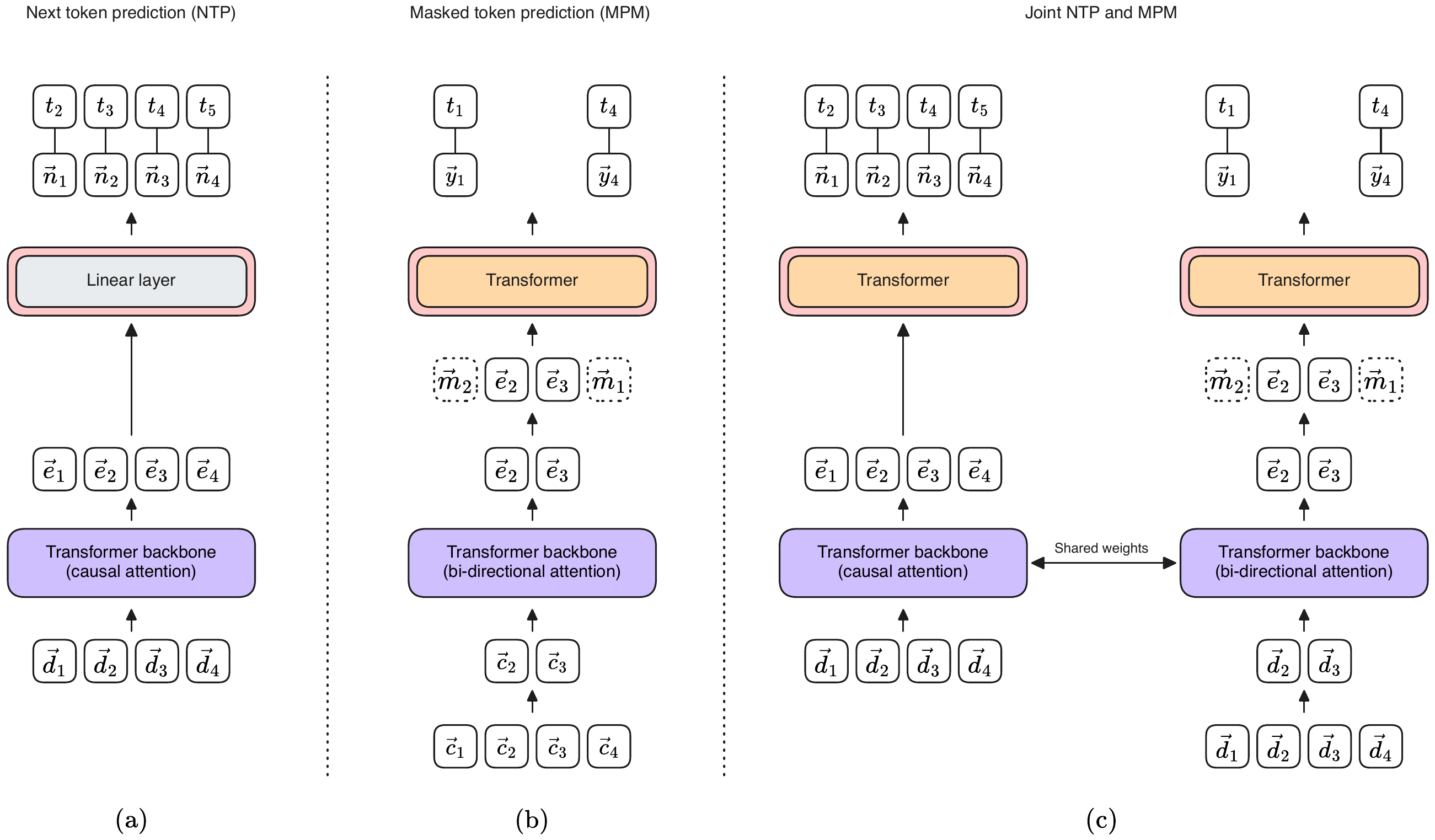}
    \caption{
        Schematic overview of the different pre-training strategies:
        (a) next token prediction (NTP),
        (b) masked token prediction (MPM)~\cite{Golling:2024abg,Leigh:2024ked},
        and (c) joint NTP and MPM.
        Token-IDs are shown as $t_i$, continuous feature vectors as $\vec{c}_i$,
        pseudo-continuous feature vectors as $\vec{d}_i$, the mask embeddings as
        $\vec{m}_i$.
        The vectors $\vec{e}_i$ represent the backbone output and $\vec{n}_i$ and
        $\vec{y}_i$ represent the next and masked token predictions, respectively.
        The particles are not $p_\mathrm{T}$-sorted, but with positional encoding
        the information of the $p_\mathrm{T}$-order within the masked subset is
        recovered: in this example, the first masked particle (position 1) is
        assumed to have smaller $p_\mathrm{T}$ than the second masked particle
        (position 4), which is why they are re-introduced as $\vec{m}_2$ and
        $\vec{m}_1$ respectively when using positional encoding.
    }
    \label{fig:sketch-generation_vs_masked-generation}
\end{figure*}

A key advantage of the original \oja{} setup is that generative pre-training produces good generative models that can be transferred to different datasets with minimal fine-tuning~\cite{Amram:2024fjg}, and earlier studies have shown that such models are directly useful for anomaly detection~\cite{Buhmann:2023acn,Mikuni:2024qsr,Mikuni:2025tar}. Beyond this immediate utility, the same pre-training was shown to significantly improve classification performance relative to training from scratch. However, the absolute classification performance obtained in this way still falls short of the current state-of-the-art. As we will show with the modifications to \oja{} detailed in the next subsection, 
this performance gap arises from the original design choice to use the discrete token-IDs as model input, as well as from the limited contextual representation learned under a NTP objective.

\subsection{Modifications}

\subsubsection{Continuous-input next token prediction}

The main advantage of using continuous input vectors is that the downstream
classification task can operate directly on the full resolution 
feature vectors $\vec{c}_i$, ensuring that the model does not suffer from
tokenization-induced loss of information in this task. With this goal of
improving downstream classification performance, we use continuous feature
vectors as model inputs while token-IDs serve only as next token prediction
targets, following the hybrid setup employed in prior works.
The general framework is shown in \Cref{fig:omnijet_alpha_workflow} (bottom row): 
during pre-training, the model acts on pseudo-continuous feature vectors,
denoted in \Cref{fig:omnijet_alpha_workflow} as $\vec{d}_i$,
which refers to the continuous representations of the tokenized inputs (i.e. the
decoded token-IDs).
During autoregressive jet generation, the decoding of token-IDs is integrated
into the autoregressive generation loop.
Note that this additional decoding step is not needed during training, 
as the whole sequence of particle tokens is fed to the model at once.

\subsubsection{Masked token prediction}

Following masked modeling approaches in natural language processing
and computer vision~\cite{devlin2019bertpretrainingdeepbidirectional,He_2022_CVPR,bao2022beit,Wang_2022_CVPR,pmlr-v162-baevski22a,Wei_2022_CVPR},
masked modeling was introduced for jet physics in the form of Masked Particle
Modeling~(MPM)~\cite{Golling:2024abg,Leigh:2024ked}. Although MPM does not produce a generative model out of the box, it has demonstrated competitive classification performance. Since the pre-training strategies of MPM and our method differ mainly in whether the model predicts masked particles or the next particle, it is a close relative to our method and we include it in our setup in order to study the effects of masked token versus next token prediction on the classification task performance.

In the MPM setup, shown in \mbox{\Cref{fig:sketch-generation_vs_masked-generation}b}, a fraction of the input particles is masked. The masked particles are removed from the input
and only the particles that survived this masking step are passed to the transformer backbone.
After the backbone, the masked particles are reinserted into the sequence in the form
of mask embeddings $\vec{m}_i$, which are trainable embeddings that are
initialized with Gaussian noise and are learned during training. Positional encoding is applied to these reinserted mask embeddings via a relation between the index $i$ and position in the $p_\mathrm{T}$-sorting: the index $i$ refers to
the position of a masked particle in the \mbox{$p_\mathrm{T}$-sorted} subset of masked
particles. That is, the i\textsuperscript{th} highest $p_\mathrm{T}$ masked particle is reintroduced
to the jet (after the backbone) as $\vec{m}_i$.
The model head is then tasked to predict the token-ID of the masked positions.
The main difference in our implementation compared to the original MPM works is that we use a significantly
smaller backbone model, which is described in more detail in
\Cref{sec:model_architecture}.

\begin{figure*}
    \centering
    \includegraphics[width=0.99\textwidth]{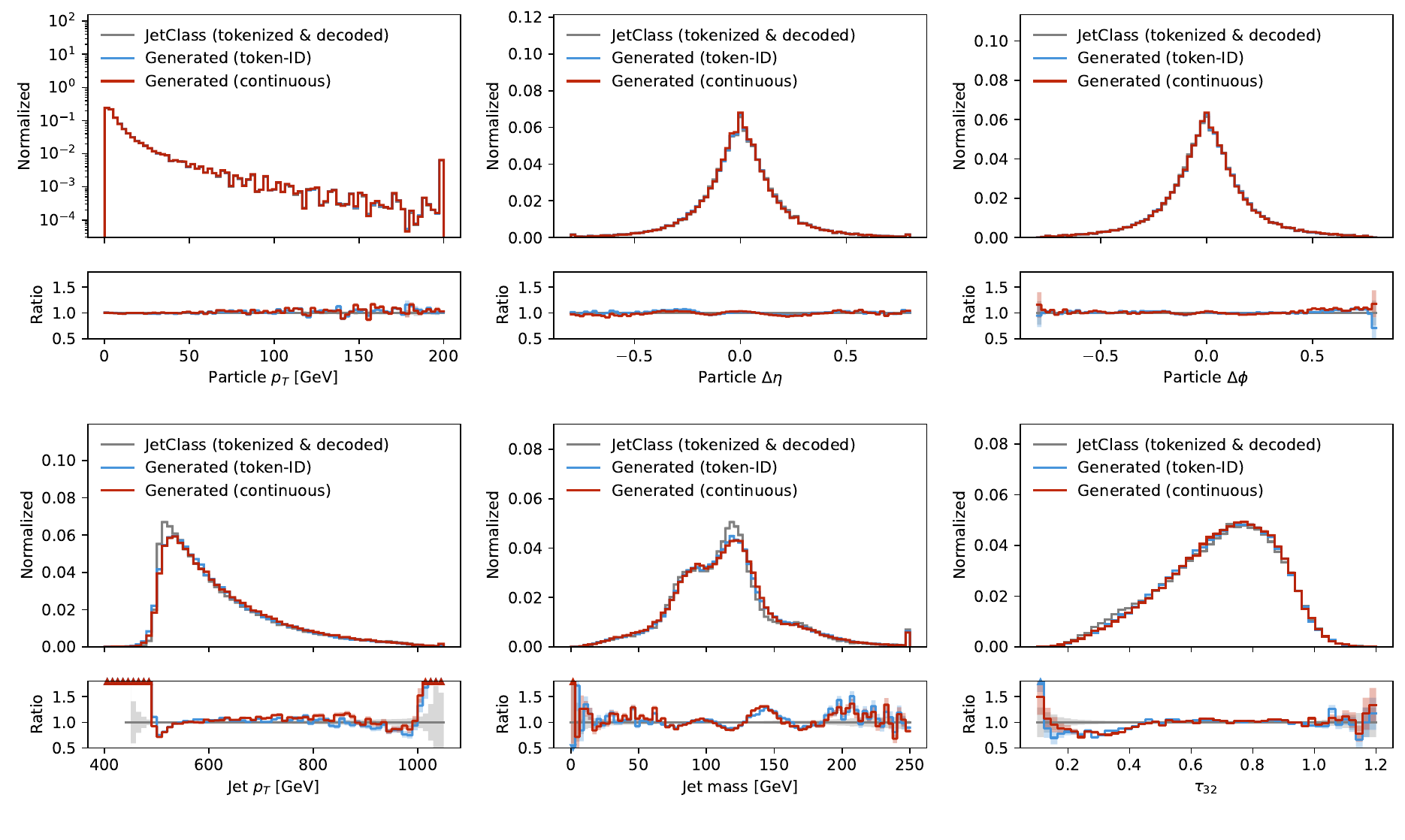}
    \caption{
        Comparison of the jets that are generated by the token-ID-input and the
        continuous-input model, as well as the jets from the JetClass dataset
        on particle-level (top row) and jet-level (bottom row).
        The first and last bins show the under- and overflow bins, respectively.
        The jets shown for the JetClass dataset are tokenized and subsequently
        decoded, representing the target of the generative model.
    }
    \label{fig:generative_results_kin_only_tokenID_vs_continuous}
\end{figure*}

A natural question is whether the benefits of MPM stem from its objective (predicting masked tokens) or some difference in the neural network architecture. The main difference is the use of bi-directional attention in MPM (required by the permutation symmetry that comes with treating the jets as point clouds), compared with the causal attention used in NTP (required by treating the jets as sequences). 
To address this, we implement a version of MPM that uses a causal attention mask in the backbone transformer as well as in the model head, in the following referred to as \emph{MPM-Causal}. This setup preserves the causal attention structure of NTP while changing only the prediction objective, allowing for a controlled comparison between the two factors. 
A masking rate of 40\% is used in all MPM(-Causal) trainings as this was
found to be the optimal masking rate in~\cite{Leigh:2024ked}.

\subsubsection{Joint next and masked token prediction}

To investigate if the advantages of both next and masked token
prediction can be combined, we implement a joint pre-training. An illustration of the joint setup is shown in \mbox{\Cref{fig:sketch-generation_vs_masked-generation}c}.
This joint pre-training is realized by having two model heads, predicting the
next token logits and the masked token logits, respectively.
The two individual loss terms are simply added without any further modification.
Two backbone forward passes are performed during pre-training.
The forward pass for the masked token prediction is performed the same way
as for pure MPM, i.e. with bi-directional attention and with positional encoding.
For the next token prediction all jet constituents are used as input and the causal
mask is applied. Furthermore, so as to not force the NTP backbone to align its output
too strongly with the generative task, we use a transformer in the NTP head in addition to the simple unembedding layer that the original \oja{} was trained with. This allows the model to
further modify the backbone representation with causal attention before
projecting the representation to next token logits.

\section{Results}
\label{sec:results}

We pre-train the backbone on the whole 100\,M-jet training set of the JetClass
dataset for each pre-training strategy.
Afterwards, we evaluate the quality of the generative model (in cases where the
pre-training involves NTP) and quantify the effectiveness of the different
pre-training strategies by fine-tuning the pre-trained models on jet
classification.
Unless otherwise specified we restrict the set of input features to the kinematic
features (particle $p_\mathrm{T}$, $\Delta\phi$ and $\Delta\eta$),
as the other features included in the JetClass dataset are not available in the top tagging dataset\footnote{We investigate the effect of extending the feature set to include the particle-ID and trajectory displacement information of the jet constituents for a model pre-trained with NTP using continuous inputs, and observe, as expected, a significant improvement in the classification capabilities. This study is shown in \Cref{sec:appendix_extended_feature_set}.}.

\subsection{Token-ID input vs. continuous input}
\label{subsec:tokenID_vs_continuous}

We first verify that the change of input structure from \mbox{token-IDs} to
continuous feature vectors can be accomplished without sacrificing the
performance of the corresponding generative model.
The jet and particle-level distributions obtained from generated jets of the two different
input types is compared to their target distributions in 
\Cref{fig:generative_results_kin_only_tokenID_vs_continuous}.
For both models the particle-level distributions of the generated jets show very
good agreement with the jets from the JetClass dataset and small deviations from
the target distribution on jet-level features.
The change in input structure (continuous feature vectors instead of token-IDs)
thus doesn't harm the performance of the generative model.

\begin{figure*}
    \centering
    \subfloat[]{
        \includegraphics[height=0.33\linewidth]{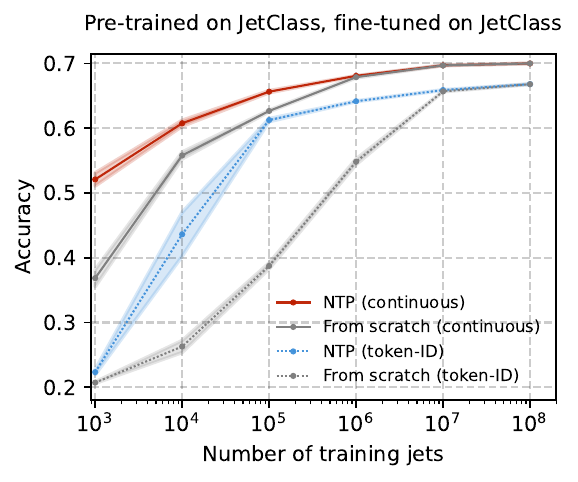}
    }\hspace{1cm}
    \subfloat[]{
        \includegraphics[height=0.33\linewidth]{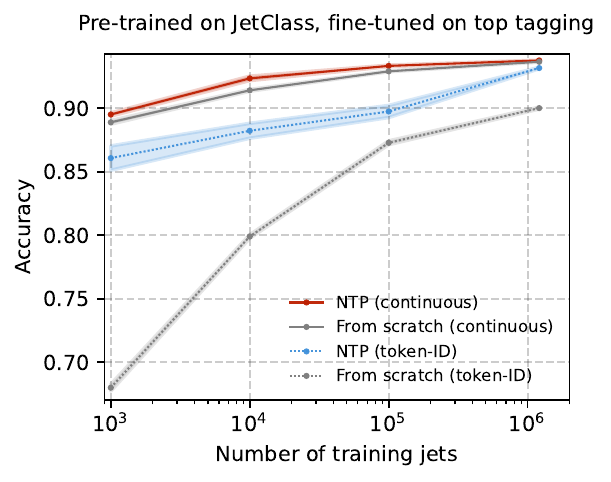}
    }
    \caption{
        Comparison of token-ID input vs. continuous input classification
        performance: (a) for multi-class classification performance on the JetClass
        dataset (all 10 jet types, in-distribution transfer learning) and (b)
        for binary classification performance on the top tagging
        (out-of-distribution transfer learning)
        as a function of the number of training jets.
    }
    \label{fig:classification_kin_tokenized_vs_continuous}
\end{figure*}

\begin{figure*}
    \centering
    \subfloat[]{
        \includegraphics[width=0.4\textwidth]{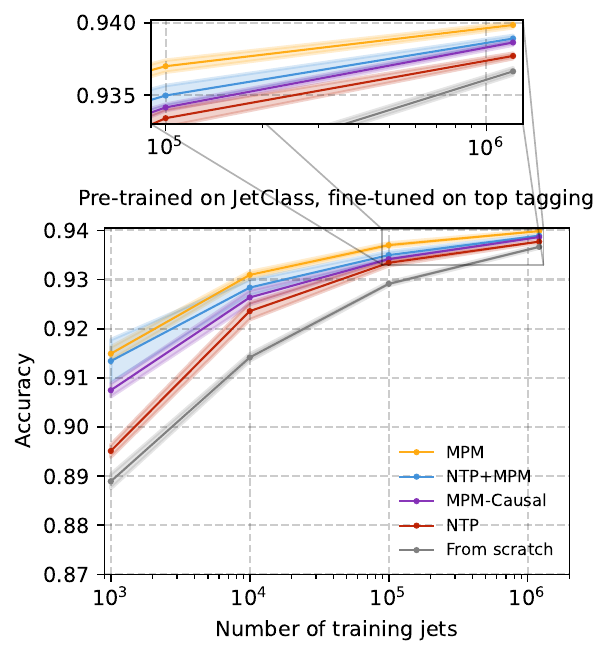}
    }\hspace{1cm}
    \subfloat[]{
        \includegraphics[width=0.4\textwidth]{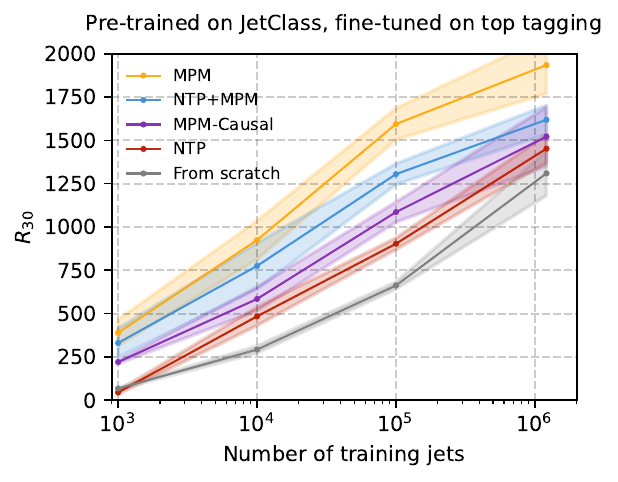}
    }
    \caption{
        Classification performance on the top tagging dataset with different
        pre-training strategies: (a) the classifier accuracy and (b) the background ($q/g$-jet) rejection at 30\,\% signal (top-jet) efficiency. All models use continuous feature vectors
        as input.
    }
    \label{fig:landscape_different_pretraining_strategies}
\end{figure*}

The positive effect of the continuous input on the classification performance
is shown in
\Cref{fig:classification_kin_tokenized_vs_continuous}a for in-distribution and in \Cref{fig:classification_kin_tokenized_vs_continuous}b for out-of-distribution transfer learning.
For in-distribution transfer learning, we fine-tune the model on multi-class
classification (10 classes) on the JetClass dataset with training dataset sizes
between 1\,k and 100\,M jets.
As expected, the \emph{from scratch} model, i.e.\
the baseline model without any pre-training, performs
significantly better with continuous input features than with token-ID inputs for all dataset sizes. This demonstrates that the tokenization alone can significantly harm the classification performance. 

The pre-trained models improve the performance compared to their respective from-scratch models in both the token-ID and the continuous case.
In the \mbox{token-ID} input case, the pre-trained model outperforms the from-scratch
model for all dataset sizes except for the largest dataset size.
In the largest dataset size regime, at 100\,M jets, the from-scratch model
catches up with the fine-tuned model, which indicates that the model has seen enough
training data to learn the relevant features from scratch and using the
pre-trained weights as initialization does not provide a benefit anymore. Meanwhile, in the continuous input case,
we still observe a performance gain from generative pre-training for small
dataset sizes, while the from-scratch and fine-tuned model converge to a similar
performance for training dataset sizes of 1\,M jets or more.

For the out-of-distribution transfer learning we fine-tune the model on the
binary classification task on the top tagging dataset. %
The dataset range for this study is between 1k and 1.2\,M jets, with
the latter being the full training dataset size of this dataset.
We see very similar trends as for in-distribution transfer learning, with
a clear performance gain when using continuous input features instead of
token-ID input features.
In this case, %
generative pre-training offers an improvement over the from-scratch baseline for
the whole dataset range, both for the token-ID input and the continuous input
case.
While the continuous input classifier presented here is still outperformed
by jet taggers that pre-train in a supervised manner directly on a classification task, like ParT~\cite{Qu:2022mxj},
L-GATr~\cite{Spinner:2024hjm,Brehmer:2024yqw} or
OmniLearned~\cite{Bhimji:2025isp}, this is expected as these studies, in
contrast to state of the art models, do not make use of either physics inspired
interaction features, class labels, or Lorentz-equivariant model architectures.

\subsection{Comparison of different pre-training objectives}

For the comparison of different pre-training strategies (NTP, MPM and joint), we pre-train with continuous inputs on the JetClass dataset with the kinematic-only feature set and fine-tune on the top
tagging dataset. 
The results of this comparison are shown in
\Cref{fig:landscape_different_pretraining_strategies}.
All investigated pre-training strategies lead to an improvement over the
baseline for the whole range of training dataset sizes, except for the
$R_{30}$ metric\footnote{
    The $R_{30}$ metric is the background rejection at 30\,\% signal efficiency.
}
at a training dataset size of 1000 jets where the fine-tuned NTP model
performs slightly worse than the from-scratch baseline.
This could be due to the fact that the difference between the pre-training
task (next token prediction) and the target task (classification) is
too large, with the model not being able to adjust to the target task with
only 1000 training jets during fine-tuning. Additional evidence for this interpretation is provided in Appendix~\ref{sec:fixed_backbone}, where fine-tuning with a fixed backbone shows that NTP-pretrained representations are substantially less aligned with classification than those obtained from MPM or joint pre-training.

\begin{figure*}
    \centering
    \includegraphics[width=0.9\textwidth]{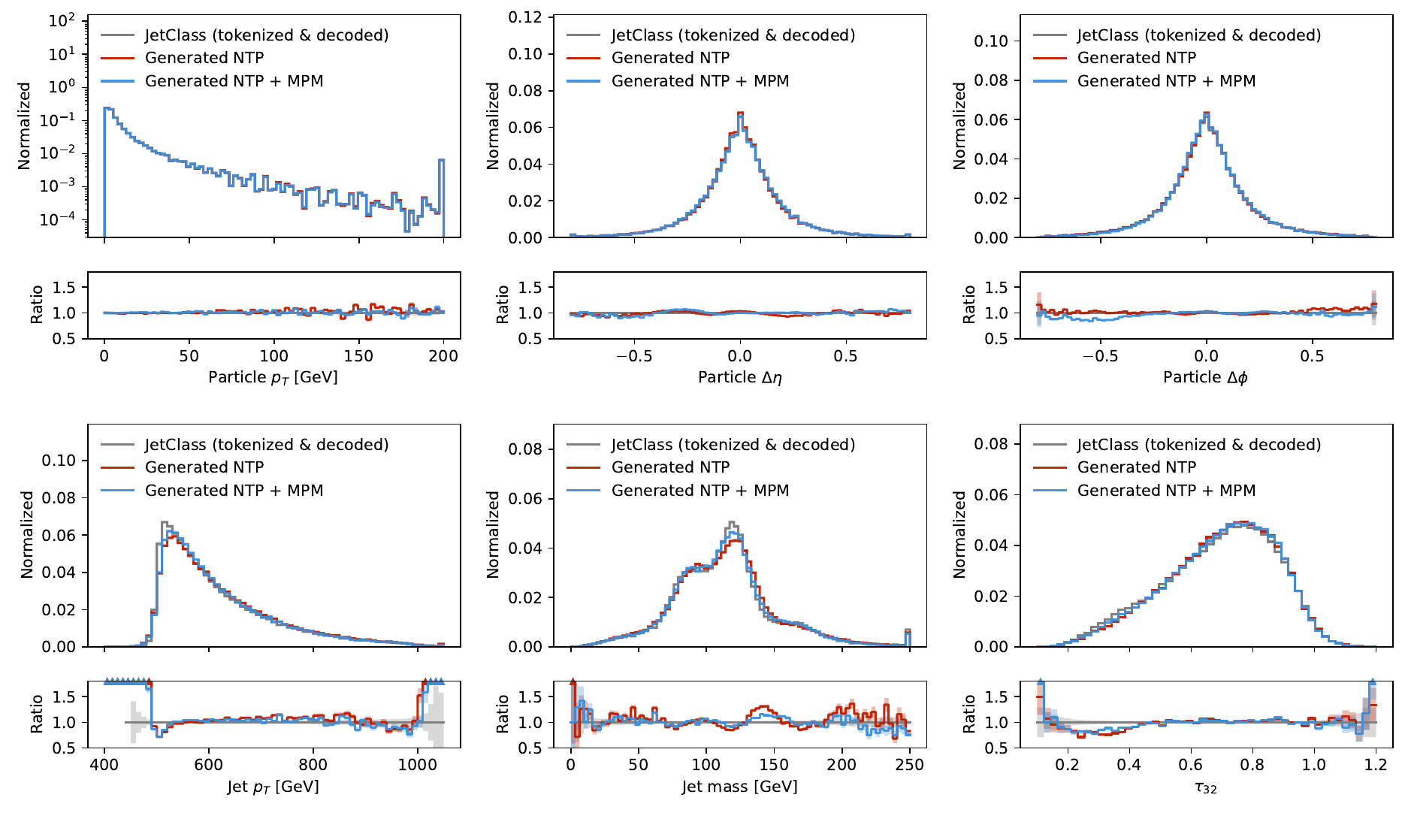}
    \caption{
        Comparison of the generative models trained with NTP only vs. joint NTP
        and MPM. The jets shown for the JetClass dataset are tokenized and
        subsequently decoded, representing the target of the generative model.
    }
    \label{fig:generative_results_NTP_vs_joint_pretraining}
\end{figure*}

While pre-training with NTP leads to a significant
improvement over the from-scratch baseline, it is outperformed by the other
pre-training strategies with the best performance being achieved by MPM.
The \mbox{MPM-Causal} setup (i.e.\ predicting masked tokens with causal
attention) performs similarly to the next token prediction setup.
This indicates that the classification performance gains seen in MPM are not
solely due to the masked prediction objective itself. The different attention
mechanisms (bidirectional vs.\ causal) also plays an important role. 
Apparently, bi-directional attention allows the model to develop a contextual understanding in a better way than causal attention does. 

The joint pre-training clearly outperforms the standard NTP pre-training and
results in a classification performance close to that of pure MPM.
At the same time, as shown in \Cref{fig:generative_results_NTP_vs_joint_pretraining},
we observe that the generative capabilities of our model are not reduced by
the joint training on NTP and MPM, with both models showing very similar
agreement with the target distribution, both on particle level and on
jet level.

\section{Conclusion}
\label{sec:conclusion}

When transplanting the methods of transformer-based foundation models from their origins in the modeling of language to their new home in solving problems in jet physics, their assumptions need to be revisited.
For example, while the concept of tokens naturally fits data that can be seen as composed from discrete elements --- such as words or word fragments --- another representation might suit the typically continuous, high resolution read-outs of modern particle detectors better. Moreover, it is not yet clear which pre-training strategy or strategies best fit the specific downstream tasks that are of interest to particle physicists. %

In this work, we have investigated these two questions: the effect of tokenizing inputs, and of setting useful pre-training tasks.
We first found that a hybrid approach with respect to tokenization can combine the best of both worlds. While the learning target for the generative pre-training step remains tokenized, it uses continuous feature vectors with limited resolution (obtained by the decoder of the VQ-VAE that created the tokens in the first place) as input. This allows the model to keep its strong generative performance, while not restricting the input format to be tokenized. 
The model can then be fine-tuned directly to classify using full-resolution continuous inputs, leading to drastic improvement of classification performance as it is no longer limited to using token-IDs as input. 

Next, we turned to the fairly subtle difference between predicting subsequent tokens (next token prediction, NTP) and predicting randomly masked-out tokens (masked particle modeling, MPM).
Naively, the difference seems marginal. However, empirically, a non-trivial difference in performance can be observed.
To better understand this behavior, we introduced the causally-masked version of MPM, allowing particles to only attend to their predecessors and thereby blocking the bi-directional attention. As this causal MPM version performs almost on-par with the generative NTP approach, one can conclude that the bi-directional attention of MPM plays a major role in aligning the learned representation for the classification task.

Finally, we investigated a joint NTP/MPM pre-training scheme and showed that it is both possible and beneficial: combining both objectives improves on the downstream classification performance while maintaining the generative fidelity of pure NTP training.
Evidently, enhancing the learned representation via some injection of MPM training does not adversely affect the generative power of NTP pretraining.

As both generative and predictive tasks are crucial applications of a successful foundation model of jet physics, the hybridization options developed in this work will aid the successful implementation of such a model.
Beyond the ideas considered here, one might further investigate whether other tokenization schemes could yield performances surpassing the VQ-VAE --- especially as the number of features is further increased. 
Alternatively, one could explore whether generation could even be achieved while fully avoiding discrete tokens~\cite{tschannen2024givtgenerativeinfinitevocabularytransformers}.
Finally, these hybrid ideas could also point towards approaches that include physical priors more explicitly, while retaining the flexibility of foundation models.

\section*{Acknowledgements}

The authors thank Oz Amram, Darius Faroughy, Michael Krämer, Alexander M\"uck, and Humberto Reyes-Gonzalez for stimulating exchanges regarding foundation models. 
We would also like to thank Thorsten Buss for valuable discussions throughout
this project.
JB, AH, GK, and NM are supported by the DFG under the German Excellence Initiative --
EXC 2121  Quantum Universe – 390833306, and by PUNCH4NFDI – project number
460248186. 
IP and DS are supported by DOE grant
DOE-SC0010008.
We also acknowledge support via the SciFM consortium (05D25GU4) funded by the German Federal Ministry of Research, Technology, and Space (BMFTR) in the ErUM-Data action plan as well as the
Hamburg VISTA/VISOR --- Virtual Initiative for Science \& Technology in AI --- network.
This research was supported in part by grant NSF PHY-2309135 to the Kavli Institute for Theoretical Physics (KITP). 

For this work the HPC-cluster \mbox{Hummel-2} at University of Hamburg was used.
The cluster was funded by Deutsche Forschungsgemeinschaft (DFG, German Research
Foundation) – 498394658.
Additionally, we acknowledge support from the Maxwell computational resources at
Deutsches Elektronen-Synchrotron DESY, Hamburg, Germany.

\section*{Code}

The code for this paper is available at
\href{https://github.com/uhh-pd-ml/enhancing-ntp4jets}{github.com/uhh-pd-ml/enhancing-ntp4jets}.

\bibliography{
    literature_HEP-ML.bib,
    literature_ML.bib,
    literature_HEP.bib,
    literature_SOFTWARE.bib
}

\appendix

\section{Model architecture and training details}
\label{sec:model_architecture}

\subsection{Backbone architecture changes}

Compared to \oja{}, which uses post-norm transformer blocks based
on~\cite{Radford2018ImprovingLU}, we use pre-norm
transformer~\cite{prenormtransformer} blocks in the backbone architecture as
this choice is more common in modern transformer architectures.
Furthermore, we add \textsc{LayerScale}~\cite{Touvron_2021_ICCV} to the
architecture, but instead of initializing the per-channel weights with a small
value as commonly done, we use a value of 1 as we found this initialization
strategy to be more effective in practice in our trainings.
We also add registers~\cite{darcet2024vision} to the backbone architecture
which was found to slightly improve classification performance in initial studies.
Overall we use a smaller backbone compared to \oja{} by using
8 transformer blocks with an embedding dimension of 128 instead of 3 transformer
blocks with an embedding dimension of 256.
Moreover, this leads to a much smaller backbone compared to the MPM work, which
uses 8 transformer blocks with an embedding dimension of 512.
The smaller model size in this work was chosen as we found that a larger
backbone did not lead to better results in initial experiments.
The token-ID-input backbone has 2.6\,M parameters as opposed to
4.5\,M parameters in the original \oja{} work.
The number of parameters is further reduced in the continuous-input backbone,
which has 1.6\,M parameters.
The much larger number of parameters in the token-ID-input case is due to the
trainable token embedding layer for the token-ID input (which maps 8192
token-IDs to the embedding dimension 128 of the backbone), resulting in around
1\,M additional parameters.
Finally, the backbone model is conditioned on the number of particles, as the
continuous-input generative model was found to struggle with correct end
token prediction when not using this conditioning in early tests.
For the case where the model acts as a generative model, we fit a kernel
density estimate (KDE) on the distribution of the number of particles per jet,
from which we sample to generate the conditional information for jet
generation\footnote{
    The generated values from the KDE are clipped to be larger than zero
    and subsequently rounded to the closest integer number.
}.
We combine jet-level information (i.e. the number of particles) with the
per-particle information by first projecting both the jet-level and the
particle-level feature vectors with a linear layer to the embedding dimension of
the backbone, and then adding the jet-level embedding to each particle
embedding.
The details of the backbone architecture and hyperparameter choices are listed in
\Cref{tab:backbone_hyperparameters}. %
Due to computational constraints, no dedicated hyperparameter optimization was performed. Instead, a few configurations were tested, 
and we selected configurations that were found to lead to stable
training performance while also not requiring an unnecessarily large model.

\subsection{Token prediction head}

As in the original \oja{} work, we use a linear layer to project the
backbone output to the logits of the token-IDs when training on the NTP target only.
The token prediction head used for MPM, as well as the token prediction head used
for the NTP prediction in the joint pre-training consists of two transformer blocks
with the same embedding dimension as the backbone, followed by a linear layer
to obtain the masked and next token logits, respectively.

\subsection{Classification head}

Compared to \oja{} we adapt the classification head to feature
two class-attention blocks \cite{Touvron_2021_ICCV,Qu:2022mxj} followed
by a linear layer (instead of two linear layers with a summation over all
particles between them) to obtain a more capable classification head. 
Furthermore, we do not apply the causal mask in the backbone
in classification mode as this is an unnecessary restriction on the model's
capacity when trained for classification.
The classification head adds another 400\,k parameters to the model, leading to
a total of 2\,M trainable parameters in the continuous-input classifier, which
is similar in size to the OmniLearn~\cite{Mikuni:2024qsr,Mikuni:2025tar} or
ParT~\cite{Qu:2022mxj} models.

\subsection{VQ-VAE architecture changes}

As the hybrid setup with continuous input and token-ID prediction requires
on-the-fly decoding of token-IDs during the generation loop, the VQ-VAE has
to be trained with a causal mask in the decoder.
Additionally, we use the same transformer block architecture as in the backbone
model, instead of the
NormFormer~\cite{shleifer2021normformerimprovedtransformerpretraining}
architecture that was used in \oja{}.
The vector quantization is implemented using the
\texttt{vqtorch}~\cite{huh2023vqtorch} library.
The VQ-VAE hyperparameters are listed in \Cref{tab:vqvae_hyperparameters}.
As for the backbone, no dedicated hyperparameter tuning was performed, rather, a
few configurations were tested and found to lead to stable results.

\subsection{Training details}

All models are trained with the Ranger~\cite{Ranger} optimizer, which uses the
Lookahead~\cite{NEURIPS2019_90fd4f88} optimizer with RAdam~\cite{Liu2020On}
as the inner optimizer.
A constant learning rate of $10^{-3}$ and a batch size of 1000 is used for all
trainings except for trainings where we use 10k jets or less, for which we use a
batch size of 100.
Pre-trainings are performed for a total of 1\,M steps, with the final model
state being used for evaluation. Classification trainings are performed for up to 1\,M
steps as well, but with early stopping based on the validation loss.
Classification runs are repeated with five different random seeds and the
average and standard deviation of the five runs is reported. The model state
with the lowest validation loss is used for evaluation in classification runs.

\begin{table}[h]
    \centering
    \caption{Backbone model hyperparameters}
    \label{tab:backbone_hyperparameters}
    \vspace{0.2cm}
    \begin{tabular}{l|c}
        Hyperparameter                                  & Value                                                 \\
        \midrule
        Number of transformer blocks                    & 8                                                     \\
        Embedding dimension                             & 128                                                   \\
        \textsc{LayerScale} initialization value~~      & 1.0                                                   \\
        Optimizer                                       & ~~Ranger~\cite{NEURIPS2019_90fd4f88,Liu2020On,Ranger} \\

        Learning rate                                   & 0.001                                                 \\
        Batch size (if $N_\mathrm{train} > 10\,000$)    & 1000                                                  \\
        Batch size (if $N_\mathrm{train} \leq 10\,000$) & 100                                                   \\
    \end{tabular}
\end{table}

\begin{table}
    \centering
    \caption{VQ-VAE model hyperparameters. For further information on $\alpha$,
        $\beta$ and $\nu$ see~\cite{pmlr-v202-huh23a,huh2023vqtorch}.}
    \label{tab:vqvae_hyperparameters}
    \vspace{0.2cm}
    \begin{tabular}{l|c}
        Hyperparameter                                                      & Value                                                 \\
        \midrule
        Number of transformer blocks (encoder)                              & 4                                                     \\
        Number of transformer blocks (decoder)                              & 4                                                     \\
        Embedding dimension                                                 & 128                                                   \\
        \textsc{LayerScale}~\cite{Touvron_2021_ICCV} initialization value~~ & 1.0                                                   \\
        Optimizer                                                           & ~~Ranger~\cite{NEURIPS2019_90fd4f88,Liu2020On,Ranger} \\

        Learning rate                                                       & 0.001                                                 \\
        Batch size                                                          & 1000                                                  \\
        Codebook size                                                       & 8\,192 (kin.)                                         \\
                                                                            & 32\,768 (ext.)                                        \\
        $\alpha$                                                            & 10                                                    \\
        $\beta$                                                             & 0.9                                                   \\
        $\nu$                                                               & 1                                                     \\
        Replacement frequency                                               & 500 steps                                             \\
    \end{tabular}
\end{table}

\begin{figure*}
    \centering
    \includegraphics[width=0.99\textwidth]{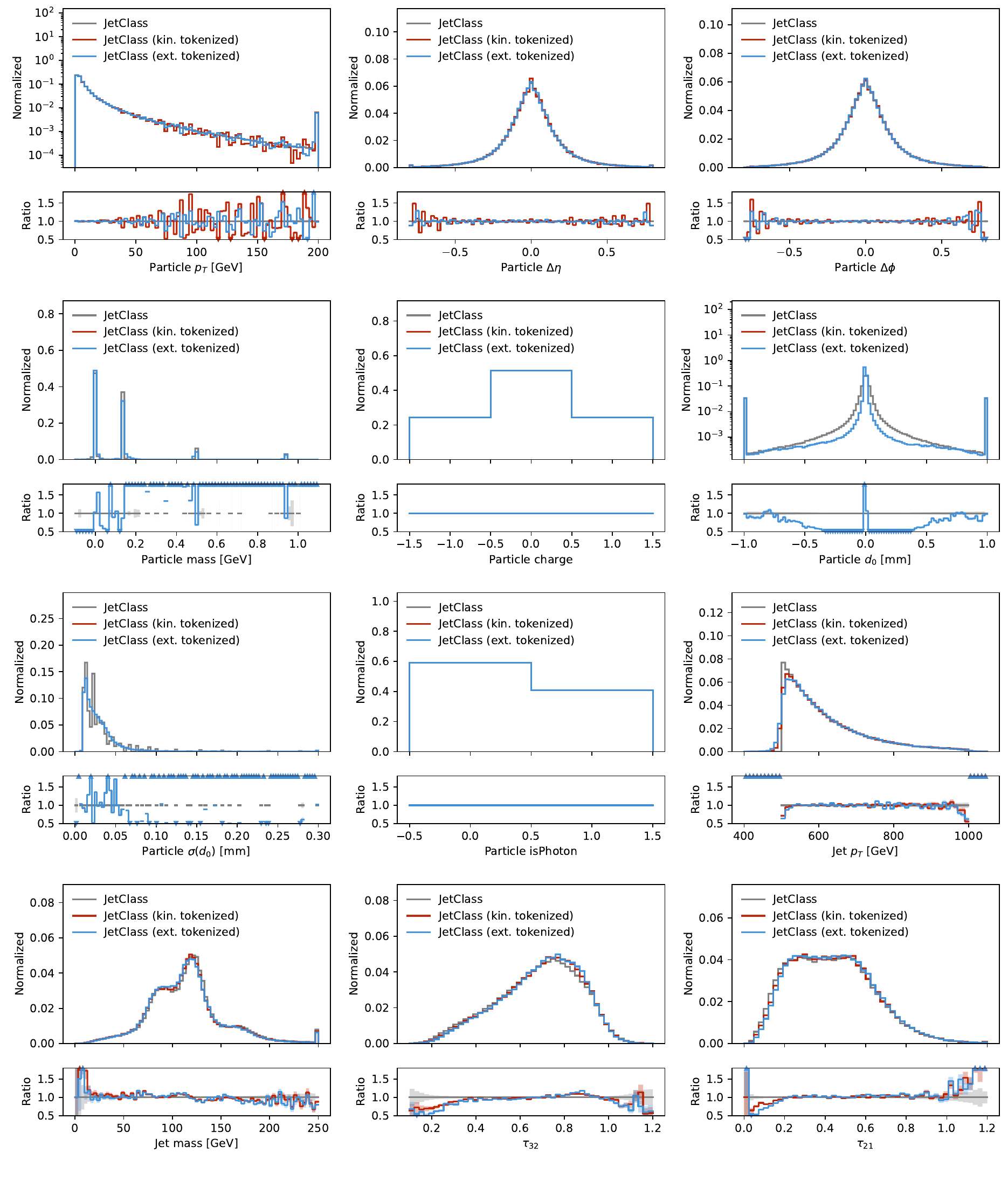}
    \caption{
        Comparison of the original JetClass dataset and the corresponding distributions
        obtained after tokenization.
        The tokenized datasets are the target for the generative models for the two
        feature sets investigated in this work: kinematics-only (kin.) and the extended
        feature set (ext.).
        Particle features related to trajectory displacement ($d_0$ and $\sigma(d_0)$) are 
        only shown for charged particles.
    }
    \label{fig:JetClass_fullres_vs_tokenres}
\end{figure*}

\section{Performance of the VQ-VAE}
\label{sec:appendix_VQVAE_performance}

A comparison of both particle-level and jet-level features of the original
JetClass dataset and the distributions obtained after tokenization (and subsequent
decoding with the VQ-VAE decoder) are shown in \Cref{fig:JetClass_fullres_vs_tokenres}.
As can be seen in the distribution of the trajectory displacement $d_0$, the VQ-VAE
struggles to accurately reproduce this distribution.
We found this feature to be very challenging to encode with a VQ-VAE, given that
it features a very sharp peak at 0 with long tails towards small and large values.
However, we leave further studies regarding high-resolution tokenization of those
features for future work.

\section{Extending the feature set}
\label{sec:appendix_extended_feature_set}

\begin{figure*}
    \centering
    \includegraphics[width=0.99\textwidth]{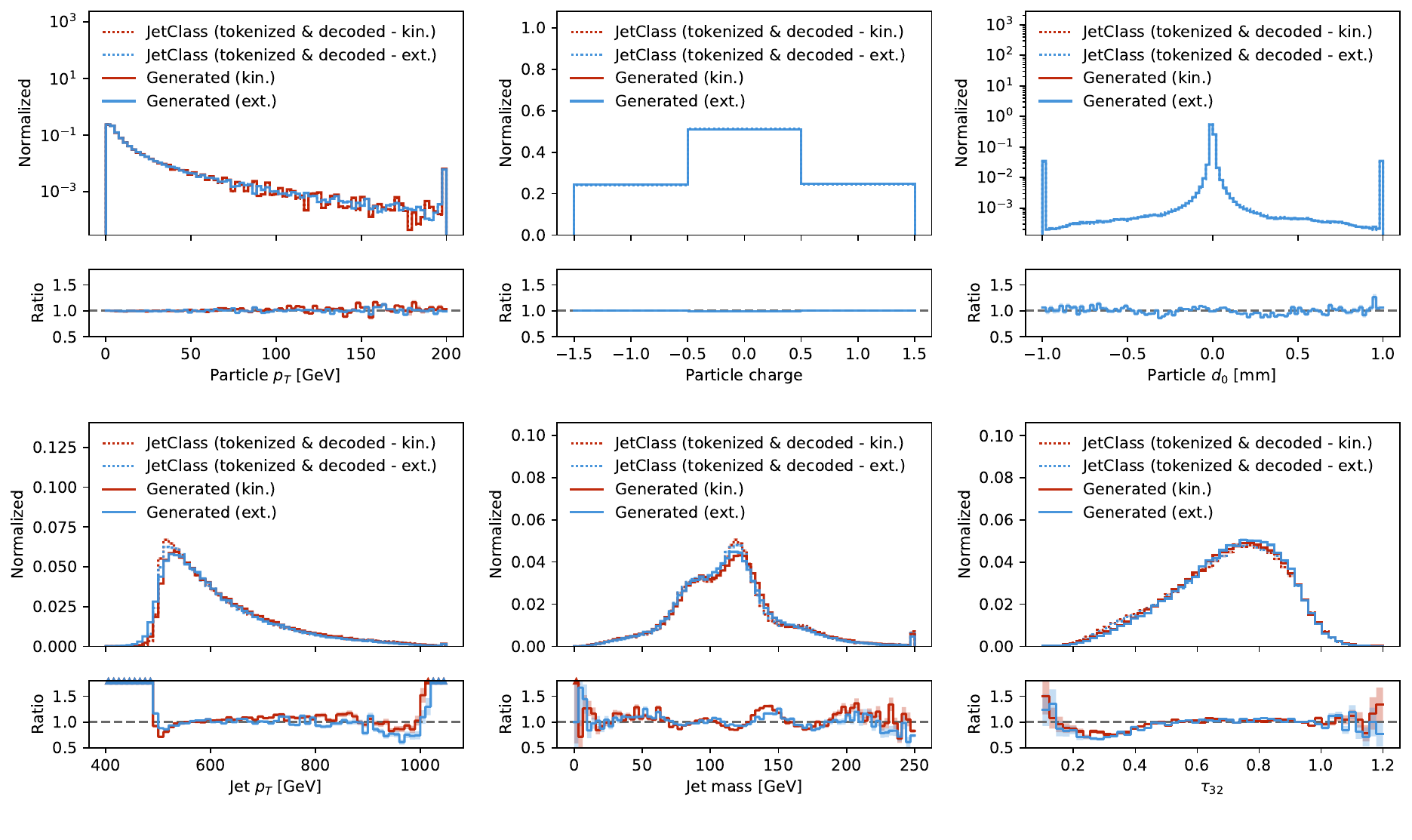}
    \caption{
        Comparison of the jets that are generated by the continuous-input model
        using the kinematics-only (kin.) and the extended feature set (ext.).
        The dotted lines show the distributions obtained from tokenization and subsequent
        decoding of the JetClass dataset with the corresponding VQ-VAE.
        The solid lines show the distributions obtained from the generative models.
        The ratio panels show the ratio between the distribution obtained from the
        generative model and the corresponding target distribution (which is shown
        in the same color in dotted linestyle).
    }
    \label{fig:generative_results_kin_vs_extended}
\end{figure*}

\begin{figure*}
    \centering
    \includegraphics[width=0.8\textwidth]{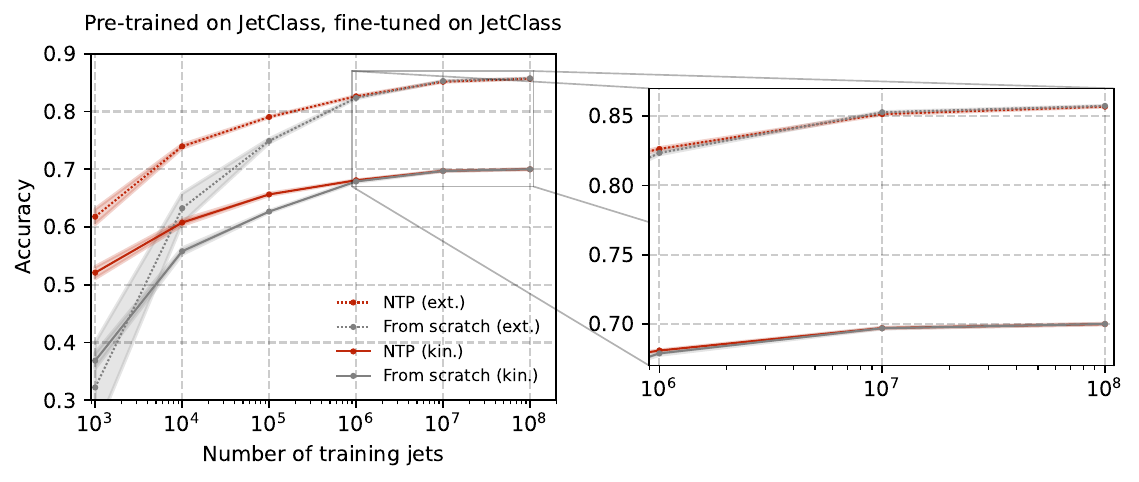}
    \caption{
        Multi-class classification performance on the JetClass dataset
        (all 10 jet types) as a function of the number of training jets
        comparing the different pre-training strategies with the corresponding
        from-scratch baselines for the kinematics-only (solid) and the extended
        feature set (dotted).
    }
    \label{fig:jetclass_continuous_kin_vs_all}
\end{figure*}

To further improve the performance and utility of our models, we extend the
input feature set of our model. We add the particle mass as well as the
particle-ID and the trajectory displacement features to the input feature set of
the model.
A new VQ-VAE is trained for the extended feature set, using a codebook size 4 times larger than for the standard feature set, and a model is subsequently
pre-trained with next token prediction.

The generative performance obtained with the two different input feature sets
is shown in \Cref{fig:generative_results_kin_vs_extended}, which shows
the particle $p_\mathrm{T}$ and two of the newly added features, the particle
charge and the transverse impact parameter $d_0$, as well as several jet level
features.
The two dotted reference lines in \Cref{fig:generative_results_kin_vs_extended} correspond
to the different VQ-VAEs used for the two different feature sets, leading
to slightly different target distributions during the generative training.

The model trained on the extended feature set shows similar performance on
observables related to the kinematics of the particles (both on particle- and
jet-level). The smoother distribution of the particle $p_\mathrm{T}$ in the
extended feature set case arises from the VQ-VAE having a larger
codebook size.
Both the particle charge distribution and the $d_0$ distribution show good agreement
with the corresponding target distribution.

By extending the input feature set, the classification performance is
drastically improved as can be seen in
\Cref{fig:jetclass_continuous_kin_vs_all}.
This overall improvement is expected as those features are known to be
useful for jet tagging.
The performance when trained on the whole JetClass dataset using the extended
feature set leads to an accuracy of 85.7\% in the baseline case whereas the
kinematics-only model reaches an accuracy of only 70\%.
Generative pre-training leads to a notable improvement in classification performance
for dataset sizes up to 1\,M jets when using the extended feature set.
For larger training dataset sizes, the performance of the pre-trained
model converges to the performance of the from-scratch model.

\section{Fine-tuning with fixed backbone representations}

\label{sec:fixed_backbone}
\begin{figure}[h]
    \centering
    \includegraphics[width=0.38\textwidth]{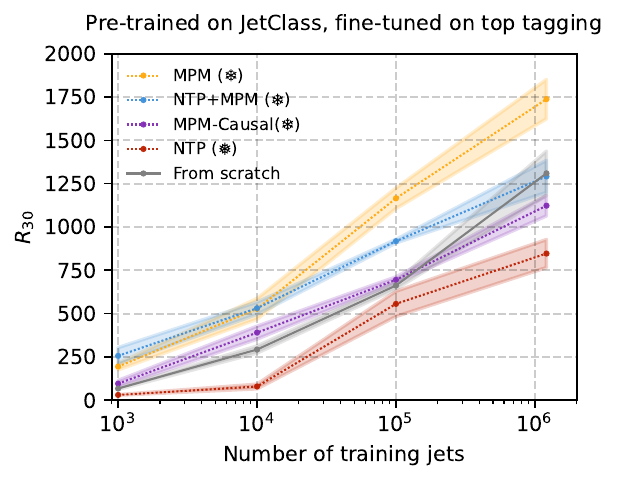}
    \caption{
        Classification performance on the top tagging dataset when fine-tuning
        with a fixed backbone.
    }
    \label{fig:frozen_finetunings_problem_description}
\end{figure}

To investigate how aligned the pre-trained backbones are with the classification task, we perform
classification fine-tuning where we only train the classifier head, shown
in \Cref{fig:frozen_finetunings_problem_description}.
While MPM and the joint pre-training perform better or on-par with the baseline
model (in which also the backbone weights are allowed to change during training)
for all dataset sizes, this is not true for MPM-Causal, which is outperformed
by the from scratch model at a dataset size of 1.2\,M training jets.
Even more notable is the behavior for the model that is pre-trained with NTP:
this model is outperformed for the whole training dataset range, meaning that
this backbone representation does not offer an advantage over the baseline when
the weights are not adapted during fine-tuning.
\footnote{
    This effect did not show up in the original \oja{} work, which
    used token-IDs as input rather than the continuous feature vector input used
    here. 
    We believe that the baseline model in that case struggles with
    classification because it also needs to ``decode'' the information contained
    in the token-IDs. The pre-training likely performs some of this decoding
    while training on the generative task, allowing it to outperform the
    baseline model. Using continuous input allows the classifier to completely
    bypass the token-IDs, making this baseline model more capable compared to
    that of ~\cite{Birk:2024knn}.
}

\begin{figure*}
    \centering
    \subfloat[bi-directional attention in classifier]{
        \includegraphics[width=0.9\textwidth]{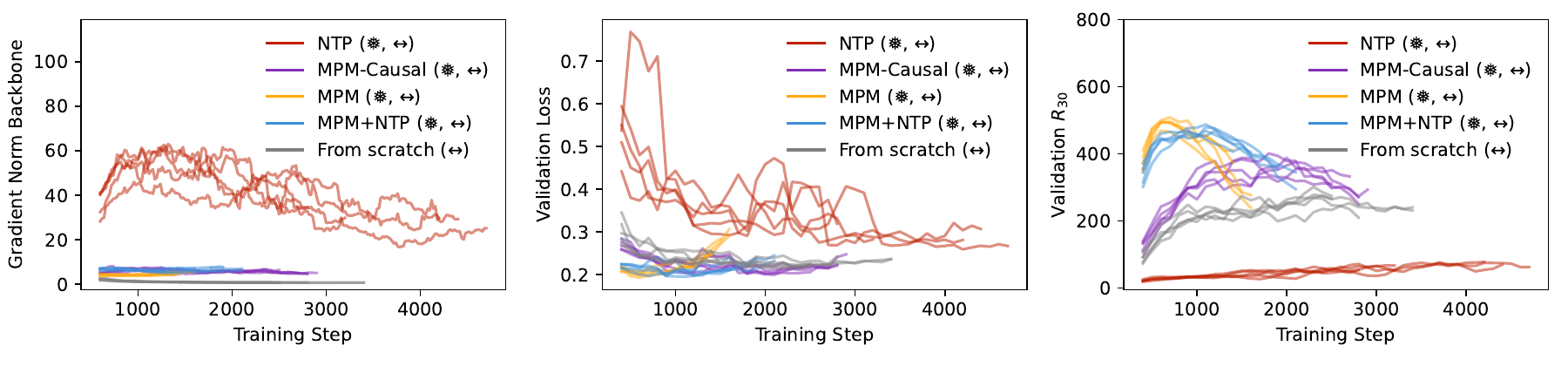}
    }\\
    \subfloat[Causal attention in classifier]{
        \includegraphics[width=0.9\textwidth]{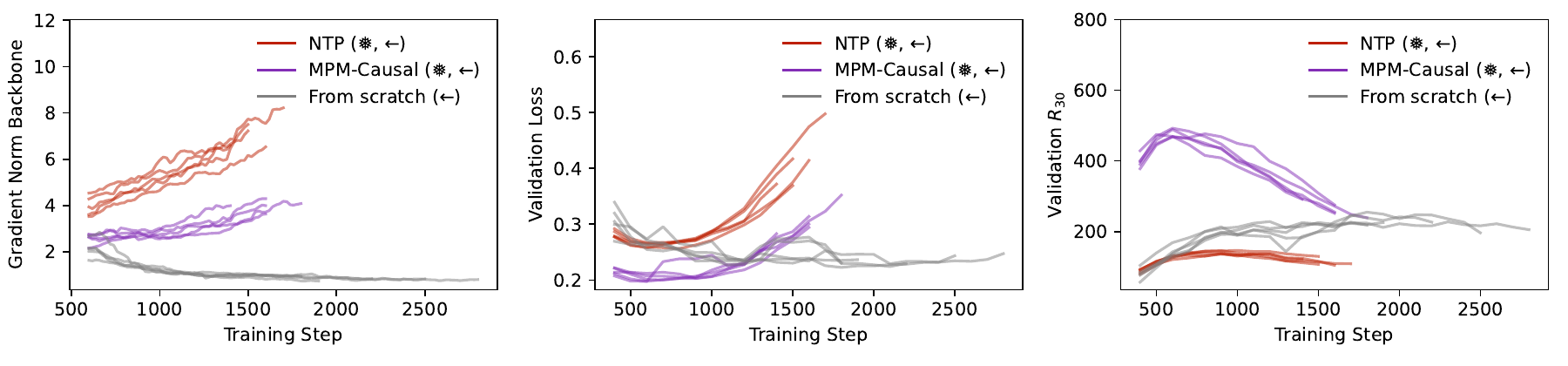}
    }
    \caption{
        Evolution of the backbone's gradient norm, as well as the
        validation loss and $R_{30}$ as a function of the training step.
        For each pre-training strategy five different runs are shown,
        corresponding to the five different runs with 10\,000 training jets in
        \Cref{fig:frozen_finetunings_problem_description}.
        The snowflake symbol indicates that the weights of the backbone are fixed. 
        The left-right arrow in the legend indicates that bi-directional attention
        is used in the backbone whereas a left arrow indicates that attention is
        only allowed to previous particles in the sequence (i.e. causal attention).
        The rolling average is shown for better readability.
        We show the average of the last 30 logged values of the gradients,
        which are logged every 20 training steps, and the average over the last
        4 epochs for the validation loss and validation $R_{30}$.
    }
    \label{fig:gradients_vs_step}
\end{figure*}

\begin{figure}[h]
    \centering
    \includegraphics[width=0.38\textwidth]{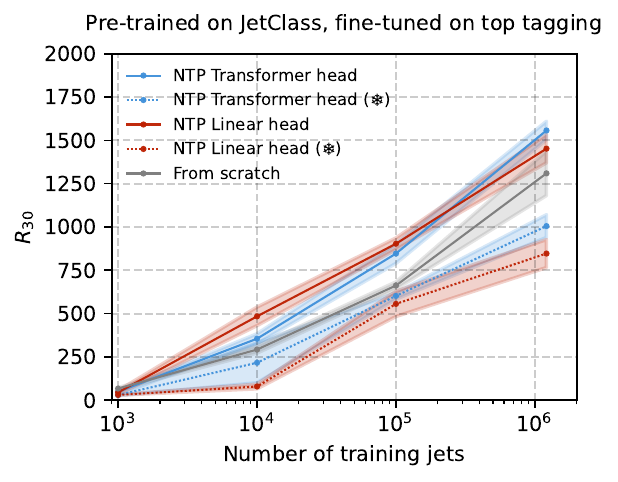}
    \caption{
        Classification performance on the top tagging dataset for NTP
        pre-trained models with two different architectures in the model head:
        \emph{Linear head} corresponds to the model head being a single linear
        layer, whereas \emph{Transformer head} corresponds to a model head with
        two transformer blocks and a linear layer.
    }
    \label{fig:gen_linear_vs_transformer_head_classification_performance}
\end{figure}

In order to explore this behavior in more detail, we study the gradients, 
validation loss and validation $R_{30}$ for the different setups. 
\Cref{fig:gradients_vs_step}a (left) shows the norm of the gradient of the backbone model
for the individual runs from \Cref{fig:frozen_finetunings_problem_description} at a dataset
size of 10\,000 jets.
The backbone model obtained from NTP pre-training shows drastically larger gradients
compared to the other methods.
MPM, MPM-Causal and NTP+MPM have larger gradients than the from-scratch baseline
(where the backbone is allowed to change) indicating that there remains
significant gradient signal to update the frozen weights. However, those
models still converges to a lower validation loss, as shown in
\Cref{fig:gradients_vs_step}a (middle).
This is not true for the NTP model, which indicates that the representations obtained
from this pre-training are not as aligned with the classification task as the
ones of the other methods, or at least is less suited for subsequent fine-tuning
to classification in a model with bi-directional attention model. 
We note that by switching from causal attention during pre-training to
bi-directional attention during fine-tuning, the fixed representation is being
used in a way that it was not trained for. And since it is fixed, it is not
able to adapt.
\Cref{fig:gradients_vs_step}b shows the corresponding behavior when using causal
attention in the classifier training.
The fine-tuning of the NTP-pre-trained model is significantly more stable, with the
loss decreasing faster and smoother.
Furthermore, the gradients show a similar behavior as the fine-tuning of the MPM-Causal
backbone. This supports the hypothesis above, that the directionality is an
important component contributing to the performance loss. However, it is not a
complete explanation as the fixed NTP backbone still lead to worse
classification performance in this scenario. 
We hypothesize that the reason for this is that the output representation of the fixed
NTP backbone will be too strongly aligned to the generative task making it more
difficult to adapt to the classification task. In order to investigate whether this can be mitigated, we compare fine-tunings with both fixed and trainable backbones for
the default case where the NTP head is just a single linear layer to fine-tunings where
the NTP head contains two transformer blocks and a subsequent linear layer like the
MPM head.
The classifier architecture is the same in all cases, only the head during pre-training
differs between the different approaches.
As shown in \Cref{fig:gen_linear_vs_transformer_head_classification_performance}, while 
the performance with the fixed backbone improves, both pre-training
architectures perform similarly in the subsequent fine-tuning when the backbone
weights are allowed to change as well.

\end{document}